\documentclass[conference]{IEEEtran}

\usepackage{graphicx}
\usepackage{float}
\usepackage{array}
\usepackage{tabularx}
\usepackage{times}
\usepackage{amsmath}
\usepackage{amssymb}
\usepackage{adjustbox}

\usepackage{verbatim}
\usepackage{moreverb}
\usepackage{alltt}
\usepackage{cite}
\usepackage{epsfig}
\usepackage{subfigure}
\usepackage{multirow}
\usepackage{slashbox}
\usepackage{indentfirst}
\usepackage{epstopdf}
\usepackage{amsfonts}
\usepackage[ruled,vlined]{algorithm2e}
\usepackage{url}

\allowdisplaybreaks[4]

\begin{document}

\title{The Performance Analysis of Coded Cache in Wireless Fading Channel}

\author{\IEEEauthorblockN{Wei Huang, Sinong Wang, Lianghui Ding, Feng Yang and Wenjun Zhang}
 Department of Electronic Engineering, \\
 Shanghai Jiao Tong University, Shanghai, China\\
Email: \{yelusaleng, snwang, lhding, yangfeng, zhangwenjun\}@sjtu.edu.cn
\vspace{-0.6cm}
}

\maketitle

\begin{abstract}

The rapid growth of data volume and the accompanying congestion problems over the wireless networks have been critical issues to content providers. A novel technique, termed as coded cache, is proposed to relieve the burden. Through creating coded-multicasting opportunities, the coded-cache scheme can provide extra performance gain over the conventional push technique that simply pre-stores contents at local caches during the network idle period. But existing works on the coded caching scheme assumed the availability of an error-free shared channel accessible by each user. This paper considers the more realistic scenario where each user may experience different link quality. In this case, the system performance would be restricted by the user with the worst channel condition. And the corresponding resource allocation schemes aimed at breaking this obstacles are developed. Specifically, we employ the coded caching scheme in time division and frequency division transmission mode and formulate the sub-optimal problems. Power and bandwidth are allocated respectively to maximum the system throughput. The simulation results show that the throughput of the technique in wireless scenario will be limited and would decrease as the number of users becomes sufficiently large.

\end{abstract}

\section{Introduction}

Wireless data traffic over the Internet and mobile networks has been growing at an enormous rate due to the explosion of available video content and proliferation of devices with increased display capabilities. According to Cisco Visual Networking Index \cite{Cisco1}, the wireless data traffic is expected to reach more than 24.3 Exabytes per month by 2019. Requests for massive data transmission over wireless systems, together with the time-varying nature of wireless connections and constraints on the available resources such as channel bandwidth and capacity, imposes significant challenges\cite{Report}.

One approach that addresses the challenges mentioned above is to bring part of the requested content closer to end users via caching\cite{cache1}\cite{cache2}. This is often referred to as the \textit{push} method. More specifically, popular contents are delivered and pre-stored in caches during relatively idle periods of wireless networks, which are retrieved later at peak time to mitigate the network congestion problem.

To further reduce the total volume of data traffic, broadcast or multicast transmissions can be incorporated into the push technique. In particular, considering that popular contents are usually requested by a large number of users, we may utilize a shared channel to deliver them via broadcast or multicast networks\cite{Networkcoding}. The performance of this technique relies heavily on the push strategy due to the gain only comes from the cache.

In contrast to pre-storing popular contents directly at caches closer to end users for future retrieval, a recently proposed implementation of the push technique relies on the idea of coded cache\cite{codedcaching1}. It can offer extra performance gain in terms of decrease in the network traffic through creating coded-multicasting opportunities by jointly coding multiple data streams at different caches\cite{codedcaching3}. With coded cache, a careful selection of content overlap across caches can ensure that multiple requests for different contents can be addressed with a single coded stream.

However, existing studies on the coded cache-based push method assumed the presence of a shared link, which is error-free and can be accessed by every user. However, in realistic wireless systems, the shared link is capacity-constrained and more importantly, it may be of different channel conditions for various users. In this case, the system performance would be restricted by the user with the worst channel quality. For illustration, consider a simple scenario where the overlapped content comes from different users with different shared channel conditions. Additional delay will be induced because it could take the user with poorer channel condition longer to transmit the required data trunk, which leads to inefficient use of wireless channel bandwidth. Therefore, an appropriate resource allocation method is needed to take full advantage of scarce bandwidth and power resources.

In this paper, we shall consider the problem of optimal resource allocation for coded cache-based push systems with a shared fading channel to address the above drawback. The study will be conducted in the context of a broadcast/multicast network and investigate how the transport mode and the number of users influence its performance. We aim to maximum the throughput with controllable power and bandwidth. To the best of our knowledge, literature on the resource allocation optimization for the coded cache-based scheme and its performance in fadding channels is still lacking.

\section{Background}

\subsection{Coded Cache}

The coded cache-based push scheme consists of two phases: the placement phase and the delivery phase. In the placement phase, each content is divided into sub-contents and part of them are pushed to users during the relatively idle periods of the wireless network. Content requests issued in the delivery phase are satisfied using the coded mutlicasting data streams.

For illustration purpose, consider the following simple example where an error-free shared link is available.

\textbf{Example 1} (Codec scheme in~\cite{coded caching2}). Suppose the server has $N=2$ contents A and B and there are $K=2$ users, each having a local cache of $MF$ bits. Each content also has $F$ bits. In the placement phase of the coded cache-based push scheme, each user randomly stores $MF/2$ bits of contents A and B, where $M<N$ \cite{codedcaching1}. Specifically, content A might be divided into four segments, denoted by $(A_{{\O}}, A_{1}, A_{2}, A_{1,2})$, where $A_{{\O}}$ represents the part of content A stored in the server, $A_{1}$ and $A_{2}$ are the parts  of content A pushed into users 1 and 2, respectively, and $A_{1,2}$ is the part of content A stored at both users. Besides, let us assume that $|A_{{\O}}|=(1-M/2)^2F\text{ bits},\ |A_{1}|=|A_{2}|=(M/2)(1-M/2)F\text{ bits},\ |A_{1,2}|=(M/2)^2F\text{ bits}$, where $|*|$ denotes the size of a data segment. Content B is divided in the same manner as content A.

In the delivery phase, suppose user 1 and user 2 request content B and A. With conventional caching scheme, the server needs to unicast $B_{{\O}}$ and $B_{2}$ to user 1 while unicasting $A_{{\O}}$ and $A_{1}$ to user 2. The total amount of data transmitted is
\begin{equation*}
2\left(\frac{M}{2}\right)\left(1-\frac{M}{2}\right)F+2\left(1-\frac{M}{2}\right)^2F=(2-M)F \text{ bits}.
\end{equation*}
On the other hand, with the coded cache, the server can satisfy the same requests by transmitting $A_{{\O}}$, $B_{{\O}}$ and $A_{2}\oplus B_{1}$ over the shared link, where $\oplus$ denotes the bit-wise XOR operation. The total traffic volume is
\begin{align*}
\left(\frac{M}{2}\right)\left(1-\frac{M}{2}\right)F+2\left(\frac{M}{2}\right)^2F&= \left(1-\frac{M}{4}\right)(2-M)F\\
&< (2-M)F \text{ bits}.
\end{align*}

The above results on data traffic reduction can be generalized to the scenario of $N>2$ contents and $K>2$ users. In this case, the realization of coded cache-based push scheme is given in the Algorithm 1. It can be seen that, in the placement phase, the local cache of each user is divided into $N$ segments with identical size, each of which stores parts of each content. In the delivery phase, the server traverses each subset of all users and transmits a coded stream to address the content requests. This procedure produces a traffic with
\begin{equation}
K\cdot (1-\frac{M}{N})\cdot\frac{N}{KM}\cdot(1-(1-M/N)^K)F\ \text{bits}.
\end{equation}
This traffic consists of two parts: the local cache gain $K\cdot (1-\frac{M}{N})$, which is produced by the uncoded caching, and the global cache gain $N/K/M\cdot(1-(1-M/N)^K)$, which is provided by the coded cache and the associated multicasting opportunities.

\begin{algorithm}[htb]
\caption{Coded Cache}
\textbf{Placement Phase}\\
\For{$(k=0; k<K; k++)$}{
\For{$(n=0; n<N; n++)$}{
user $k$ randomly prefetches $MF/N$ bits of content $n$ \;
}
}
\textbf{Delivery Phase}\\
\For{$(k=K, k>1; k--)$}{
\For{choose $k$ users from $K$ users to form a subset $U$}{
 server sends $\oplus_{k \in U}V_{k,U/\{k\}}$ to users in $U$.
}
}
\end{algorithm}

\subsection{Coded Cache-based Push in Wireless Network}

In realistic wireless networks, different users may experience different states of the shared link, which could be described using e.g., the signal-to-noise ratio (SNR) to reflect the combined effects of the channel fading and the local AWGN. This would lead to a phenomenon called \emph{multicast saturation} \cite{ms}, where the channel capacity remains almost unchanged when the number of multicast users becomes sufficiently large and continues to increase. The capacity saturation has been shown to be due to the limitation imposed by the user with the worst channel condition.

The above problem would appear in the context of coded cache-based push method in the wireless networks. This can be seen by examining Algorithm 1, where multicasting transmissions to different users are performed in the delivery phase. In the following sections, we shall first present the deployment of the coded caching scheme in the wireless network and then proceed to developing resource allocation schemes to combat the impact of different channel states at various users.

\section{Network Architecture}

\begin{figure}
  \centering
  \vspace{-0.1cm}
  \includegraphics[scale=0.4]{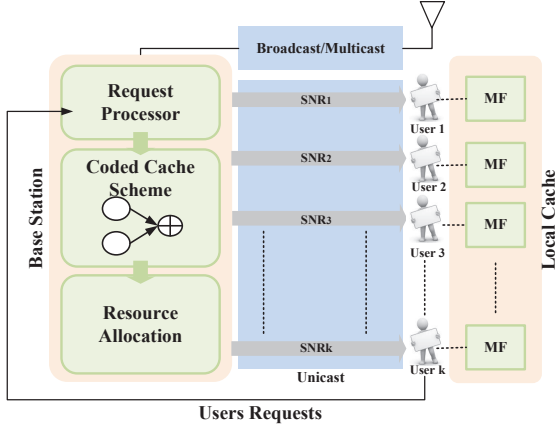}\\
  \vspace{-0.1cm}
  \caption{The network architecture}\label{Fig1}
  \vspace{-0.1cm}
\end{figure}

Consider a wireless network shown in Fig. \ref{Fig1}. Associated notations are listed in Table I. The network in consideration consists of a wireless broadcast or multicast network with fading channels, which are accessible by $K$ users in the covered area. Each user has a local cache of $MF$ bits. The AWGN at each user has a PSD $n_{k}/2$, where $k=1,2,...,K$. The server of the wireless system has $N$ contents and it is connected to the broadcasting system through wired backhaul. Each content has the identical size of $F$ bits. The coded cache-based push method is adopted and we assume that each user's local cache has been configured in the placement phase using Algorithm 1.

\begin{table}[t]
\caption{Main notations} \vspace{-0.2cm}
\begin{center}
\begin{tabular}{| >{\bfseries }l | p{7 cm } |}
\hline\hline
 $N$ & The number of contents in the server.\\
  \hline
  $K$ & The number of users in the system.\\
  \hline
  $F$ & The size of each content.\\
  \hline
  $V_{i}$ & Content $i$.\\
  \hline
  $n_{i}$ & PSD of user i.\\
  \hline
  $M$ & The number of contents can be prefetched by each user.\\
  \hline
  $d_{k}$ & The index of the content requested by user $k$.\\
  \hline
  $X_{ij}^{k}$ & The indictor represents the resource allocation for $k$th transmission in $i$th time slot and $j$th subcarrier.\\
  \hline
  $B_{u}$ & The bandwidth of each subcarrier.\\
  \hline
  $T_{u}$ & The time of each time slot.\\
  \hline
  $H$ & The number of subcarriers in the system.\\
  \hline
  $B_{k}(i)$ & The bandwidth allocated for $k$th transmission in $i$th time slot.\\
  \hline
  $P_{k}(i)$ & The power allocated for $k$th transmission in $i$th time slot.\\
  \hline
  $U_{k}$ & The receiver set in $k$th transmission.\\
  \hline
  $S_{k}$ & The signal size of $k$th transmission.\\
\hline \hline
\end{tabular}
\end{center}
 \label{tab:notation}
 \vspace{-0.7cm}
 \end{table}

In the delivery phase, user $k$ sends its request $d_{k}$ to the base station. The base station collects all users' requests and take the coded delivery in Algorithm 1. It can be found that there are $2^K$ coded transmissions in this phase, the $2^K-K$ of which are the multcasting transmission. The rest $K$ transmissions are unicasting transmission that unicast the unprefetched bits of each requested contents. For example, the base station unicasts $B_{{\O}}$ to user 1 and $A_{{\O}}$ to user 2 in the \textbf{Example 1}. Thus, we use the broadcasting system to accomplish first $2^K-K$ transmissions, then use the cellular system to unicast the unprefetched part of each user's request.

\section{Resource Allocation}

We shall investigate the problem of resource allocation via taking into account the fact that under the considered wireless network, each transmission may experience different channel fading and the AWGN of receiving users may have different PSDs (i.e., the received SNR of different transmissions could be varying)\cite{RS3}. For this purpose, let us denote the average transmitting power and the bandwidth of the base station as $P$ and $B$.
Besides, we adopt the simplified model in \cite{RS1}, where it is assumed that the available bandwidth is partitioned evenly into $H$ subcarriers and the transmission is time-slotted. Let $B_{u}$ and $T_u$ be the bandwidth of each subcarrier and the duration of each time slot such that $H\cdot B_u=B$.

To formulate an optimization problem for resource allocation, we define $X_{ij}^{k}$ to be a binary variable such that $X_{ij}^k=1$ if the $i$th time slot and the $j$th subcarrier are allocated to the $k$th transmission; otherwise, $X_{ij}^k=0$. The set of feasible solutions to the resource allocation problem can then be expressed as
\begin{equation}
\mathfrak{F}=\left\{\{X_{ij}^{k}\}\big|X_{ij}^{k}\in\{0,1\},\sum\limits_{k=1}^{2^K}X_{ij}^{k}=1, \forall (i,j)\right\}
\end{equation}
where the equality constraint comes from the one-to-one correspondence in the sense that any time slot and subcarrier can only be assigned to one transmission. Besides, we denote $t_{k}$ as the duration of the $k$th transmission, $B_{k}(i)$ and $P_{k}(i)$ as the bandwidth and power allocated for $k$th transmission in the $i$th time slot. Finally, let $U_{k}$ be the set of users receiving the $k$th transmission and $S_{k}$ be the number of bits in the $k$th transmission. $S_k$ can be evaluated as follows. Consider a particular bit of a certain content. The probability that it is pushed into a user's local cache is $M/N$. Consider the user subset $U_{k}$ that has $|U_{k}|$ users and is receiving the $k$th transmission. The probability that this bit is stored solely at each of those $|U_{k}|$ users is $(M/N)^{|U_{k}|}(1-M/N)^{K-|U_{k}|}$. Thus, the average signal size of the $k$th transmission is
\begin{equation}
S_{k}=F\cdot\left(1-\frac{M}{N}\right)^{K-|U_{k}|}\cdot\left(\frac{M}{N}\right)^{|U_{k}|}.
\end{equation}

With the above notations, the generic resource allocation problem for coded cache-based push scheme for wireless networks can be formulated as

\textbf{OPT:}
\begin{equation}
\min\sum\limits_{k=1}^{2^K}\omega_{k}t_{k}
\end{equation}
\begin{align}
\text{subject to}\quad &\sum\limits_{i=1}^{t_{k}}B_{k}(i)\log_{2}\left[1+\frac{P_{k}(i)}{n_{k}^{m}B_{k}(i)}\right] \geq \frac{S_{k}}{T_{u}}, \forall k\\
&n_{k}^{m}=\max\limits_{j\in U_k}\{n_{j}\}, \forall k\\
&0 \leq B_{k}(i) = \sum\limits_{j=1}^{H}X_{ij}^{k}B_{u} \leq B, \forall k,i\\
&\sum\limits_{k=1}^{2^K}B_{k}(i)\leq B,\quad\sum\limits_{k=1}^{2^K}P_{k}(i)\leq P\\
\text{variable}\quad &X_{ij}^{k}\in\mathfrak{F}
\end{align}

The objective function is the weighted sum of the transmission time $t_{k}$ for the $k$th transmission. Since the traffic volume under coded caching scheme for each transmission is fixed, maximizing the system throughput becomes equal to minimizing the total transmission time. The weighting factors $\omega_{k}$ are introduced to generalize the formulation. The constraints (5) and (6) represents that for each transmission and under the limitation from the user with the worst channel condition, the resource allocation scheme needs to allow a sufficient channel capacity in order to guarantee the delivery of that transmission. Constraints (7) and (8) correspond to that the amount of allocated bandwidth and transmitting power should not overtake their maximum allowable values. The OPT falls into a nonlinear integer Programming (NIP) model , which is NP-hard in the strong sense. In fact, a real resource allocation cannot be such optimal and should follow some rules. In the following section, we will relax this model to get a sub-optimal solution and examine the solution to the generic resource allocation problem in (4) in two modes, namely, the time-division (TD) mode and the frequency-division (FD) mode.

\subsection{Time-division}

\begin{figure}
  \centering
  \vspace{-0.3cm}
  \includegraphics[scale=0.4]{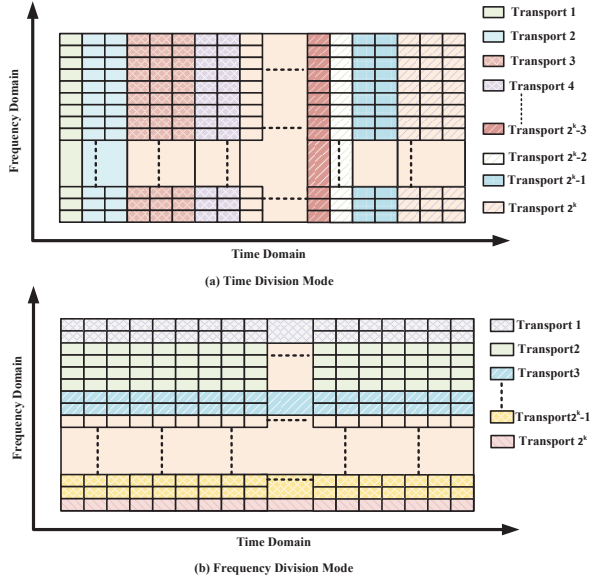}\\
  \vspace{-0.3cm}
  \caption{The resource allocation in TD transport mode and FD transport mode
   Each colour represents one transmission and each transmission gets different resource
  }\label{Fig3}
  \vspace{-0.6cm}
\end{figure}

In the time-division (TD) transport mode, the transmissions are performed in sequence but each transmission would utilize the available bandwidth in a whole. In this case, the total transmission time equals to the sum of the number of the time slots of all transmissions. Hence, $\omega_{k}=1, \forall k$ and the decision space is,
\begin{equation}
\mathfrak{F}^{T}=\left\{\{X_{ij}^{k}\}\big|X_{ij}^{k}\in\mathfrak{F},\sum\limits_{j=1}^{H}X_{ij}^{k}=H\cdot Y_{i}^{k}, \forall (i,k)\right\}
\end{equation}
where $Y_{i}^{k}$ is a binary variable such that $Y_{i}^{k}=1$ if the $i$th time slot is allocated to $k$th transmission. Putting the definition of $Y_{i}^{k}$ into (5)-(8) and replacing the decision space (2) by (10), we obtain the resource allocation problem for coded cache-based push technique under TD transport mode.

To further simplify the optimization problem, we adopt the equal power TD technique, where the transmit power $P$ and bandwidth $B$ are allocated to each transmission for a fraction $\tau$ of the total transmission time. Under the assumption that the $i$th transmission consumes $\tau_{i}\times 100$ percent of the total transmission time, the resource allocation problem can then be re-written as \\
\textbf{OPT-TD:}
\begin{equation}
\min\quad\sum\limits_{i=1}^{2^K}\frac{S_{i}}{\tau_{i}B\log_{2}\left(1+\frac{P}{n_{i}^{m}B}\right)}
\end{equation}
\begin{align}
\text{subject to}\quad&\sum\limits_{i=1}^{2^K}\tau_{i}=1, 0<\tau_{i}<1, 1\leq i \leq 2^K\\
&n_{i}^{m}=\max\limits_{j\in U_i}\{n_{i}\}, 1\leq i \leq 2^K\\
\text{variable}\quad&\tau_{i}
\end{align}

In the \textbf{OPT-TD} model, the objective function is the total transmission time to satisfy all users' requests. The principle of the above optimization framework is illustrated in Fig. \ref{Fig3}(a). Applying the \emph{Cauchy-Schwarz Inequality}, we can obtain the optimal solutions which are given by
\begin{equation}
\tau_{i}^{*}=\sqrt{\frac{S_{i}}{\log_{2}\left(1+\frac{P}{n_{i}^{m}B}\right)}}\big/\sum\limits_{j=1}^{2^K}\sqrt{\frac{S_{j}}{\log_{2}\left(1+\frac{P}{n_{j}^{m}B}\right)}}.
\end{equation}

\subsection{Frequency-division}

When it comes to frequency-division(FD) mode, since the time slot is fully allocated to each transmission, each transmission is operated in the parallel manner. Therefore, the total transmission time is determined by the transmission that has the longest time and the $\omega_{k}=E[t_{k}]^{\infty}, \forall k$ and the decision space is,
\begin{equation}
\mathfrak{F}^{F}=\left\{\{X_{ij}^{k}\}\big|X_{ij}^{k}\in\mathfrak{F},\sum\limits_{i=1}^{t_{k}}X_{ij}^{k}=t_{k}\cdot Z_{j}^{k}, \forall (j,k)\right\}
\end{equation}
where $Z_{j}^{k}$ is a binary variable such that $Z_{j}^{k}=1$ if the $j$th subcarrier is allocated to the $k$th transmission. $Z_{j}^{k}$ satisfies $\sum_{k=1}^{2^K}Z_{j}^{k}=1$.Putting the definition of $Z_{j}^{k}$ into (5)-(8) and substituting the decision space (2) using (11), the optimization problem for resource allocation in the FD mode is established.

Similar to the TD case, under the assumption that the base station allocates $P_{i}$ of its total power $P$ and $B_{i}$ of its total bandwidth $B$ to the $i$th transmission, we can write the resource allocation problem for coded cache-based push method in wireless network as \\

\textbf{OPT-FD:}
\begin{equation}
\min\quad\max\limits_{1\leq i \leq 2^K}\left\{\frac{S_{i}}{B_i\log_{2}\left(1+\frac{P_{i}}{n_{i}^{m}B_i}\right)}\right\}
\end{equation}
\begin{align}
\text{subject to}\quad&\sum\limits_{i=1}^{2^K}P_{i}=P, 0<P_{i}<P\\
&\sum\limits_{i=1}^{2^K}B_{i}=B, 0<B_{i}<B\\
&n_{i}^{m}=\max\limits_{j\in U_i}\{n_{i}\}, 1\leq i \leq 2^K\\
\text{variable}\quad&B_{i}, P_{i}
\end{align}

In the \textbf{OPT-FD} model, the objective function is the total transmission time to satisfy all requests. Fig.\ref{Fig3}(b)  shows the principle of the resource allocation problem in the FD mode.  The solutions to the \textbf{OPT-FD} problem can be found via the application standard nonlinear programming algorithms.

\section{Numerical Results}

In this section, the proposed resource allocation scheme including TD and FD, are simulated in a broadcast network. For comparison, we use the traditional uncoded cache method as the baseline scheme.

\textbf{Baseline scheme:} The system only consists of a cellular network, which unicasts the unrepfetched content into each user. And we consider the same resource optimization model to get the maximum system throughput. For simplification, we denote it as '$B$' and the coded caching scheme as '$C$'.

The simulations operated under the following assumptions. The users are uniformly distributed in a cell with a large-scale path loss of 2. The broadcast radius is $5$km. The number\footnote{In simulations, we have difficulty in obtaining the optimal solution for large values of $K$, due to the $2^K$ transmissions of coded cache scheme} of users $K$ is between 2 to 128. The channel is frequency selective Ricean Fading channel. The noise variance $n/2=1$.

\subsection{Impact of cache size}

From the previous theoretical analysis~\cite{codedcaching1}\cite{codedcaching2}, the traffic volume can be reduced due to the prefetching of local cache, and the coded caching scheme introduce a global cache gain, which shows a huge traffic reducing gain compared to the baseline scheme. However, in the wireless fading channel, one of the fundamental questions is how the system throughput scales as the size of local cache. Does there exist a global cache gain due to the coding? Based on above resource allocation model, Fig.\ref{Fig4} and Fig.\ref{Fig5} shows the comparison of coded caching scheme and baseline scheme when the cache size increases. We have following observations:

\begin{figure}[htb]
\centering
\vspace{-0.3cm}
\includegraphics[width=0.42\textwidth]{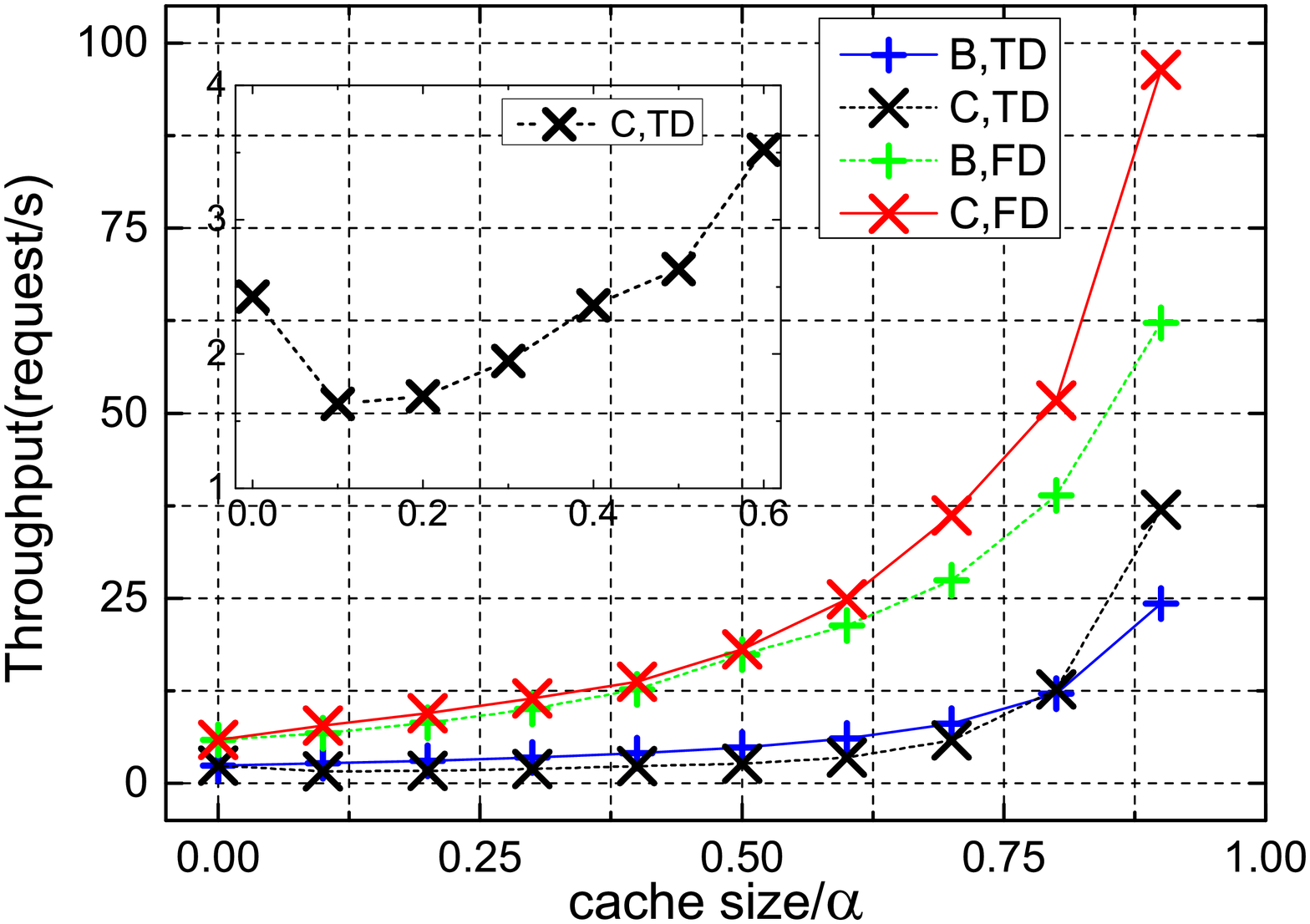}
\vspace{-0.4cm}
\caption{The relationship of system throughput and cache size.}\label{Fig4}
\label{fig:usernum}
\vspace{-0.3cm}
\end{figure}

The system throughput increases as the cache size increases under above two schemes and two kinds of resource allocation strategies, except the coded caching scheme under time division. And the performance of both schemes under FD mode are better than TD mode.

We plot the throughput gain of the coded caching scheme compared to the baseline scheme in Fig.\ref{Fig5}. Under FD mode, coded caching scheme shows an approximate constant throughput improving gain of $1-1.5\times$. Under TD mode, it first decreases, then increases up to $1.5\times$. For comparison, we also plot the traffic gain of coded cache scheme, which is much larger than the throughput gain. These results show that, in the wireless communication setup, the gain of coded cache is limited. The main reason is that, the system throughput is bounded by both traffic volume and channel rate. Although the traffic volume is reduced a lot due to the coded multicast, the multicast capacity is restricted by the user that has the worst channel gain.

\begin{figure}[htb]
\centering
\vspace{-0.3cm}
\includegraphics[width=0.42\textwidth]{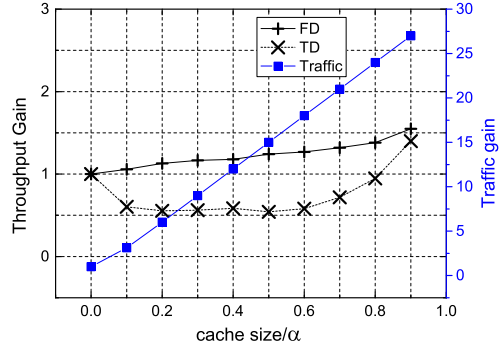}
\vspace{-0.3cm}
\caption{System Throughput and traffic gain versus cache size.}\label{Fig5}
\label{fig:usernum}
\vspace{-0.6cm}
\end{figure}

As shown in the Fig.\ref{Fig4}, the system throughput of coded caching scheme under time division shows a unique single valley manner. Under small cache size $\alpha<0.1$, the system throughput is decreasing, while under large cache size $\alpha>0.1$, the system throughput is increasing. The main reason behind this trend is that \emph{the traffic reduce due to the cache size increasing cannot improve the system throughput under TD}. Here we show a possible explanation. From the theoretical analysis in Section V.A, the optimal solution under TD strategy equals to a weighted sum of all $|S_{i}|$ and contains a root square form, which will weaken the traffic volume reducing effect of $|S_{i}|$ by the extraction of it and weighed summation.

\begin{figure*}[htb]
\centering
\includegraphics[width=0.99\textwidth]{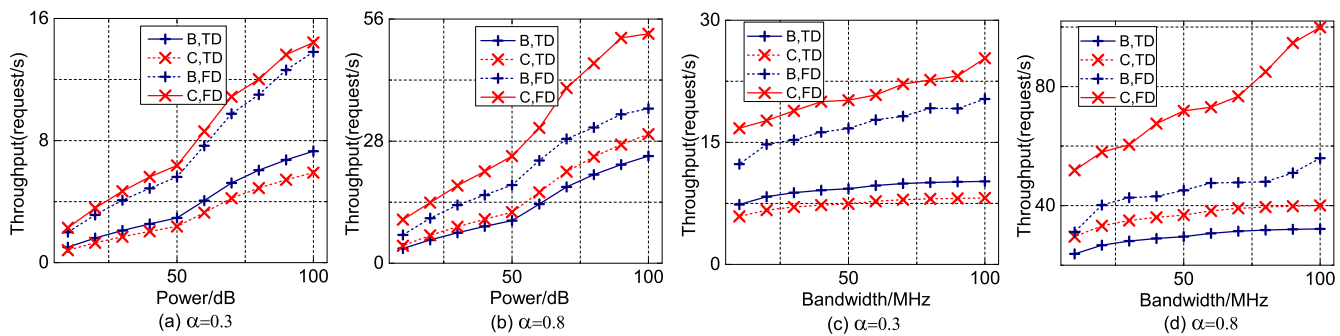}
\vspace{-8.5cm}
\caption{System Throughput versus Power and Bandwidth under two kinds of cache size.}\label{Fig6}
\vspace{-0.4cm}
\end{figure*}

\subsection{Impact of system parameter}

Then we show the impact of the network resources, such as the system bandwidth $B$ and the system power $P$, on the network performance. Based on the prior results we get, both coded caching scheme and base scheme show different performances with different cache sizes. Thus, we conduct the following simulation under two kinds of cache sizes.

\subsubsection{The effect of $P$}

Fig.\ref{Fig6}(a) and Fig.\ref{Fig6}(b) plots the system throughput versus system power under two kinds of cache sizes. It can be seen that the system throughput increases when the system power increases, and the gap between above schemes under two strategies is constant. And coded caching scheme performs a bit better than conventional scheme in FD modes.

\subsubsection{The effect of $B$}

Fig.\ref{Fig6}(c) and Fig.\ref{Fig6}(d) plots the system throughput versus system bandwidth under two kinds of cache size. It can be seen that the throughput also increases with bandwidth. However, when the cache size is small, the performance of coded caching scheme under TD mode is even worse than conventional one. In contrast, when it comes to the FD mode, given a larger cache size, coded caching scheme with resource allocation is much more better than the conventional scheme.

\subsection{Impact of user channel fading}

From the previous analysis, we show that, in the wireless fading environments, the coded caching scheme almost has a constant gain compared to the baseline scheme, which is caused by the wireless channel fading phenomenon. Moreover, we investigate the relationship between system throughput of both schemes and the degree of channel fading. Since a multicast system saturates the capacity when the number of users increases, it is necessary to investigate how the system throughput scales as the the number of users. In Fig.~\ref{fig:usernum}, we plot the system throughput versus the number of users with $M/N=\{0.3,0.8\}$. For comparison, we do not consider the outage.

\begin{figure}[htb]
\centering
\includegraphics[width=0.5\textwidth]{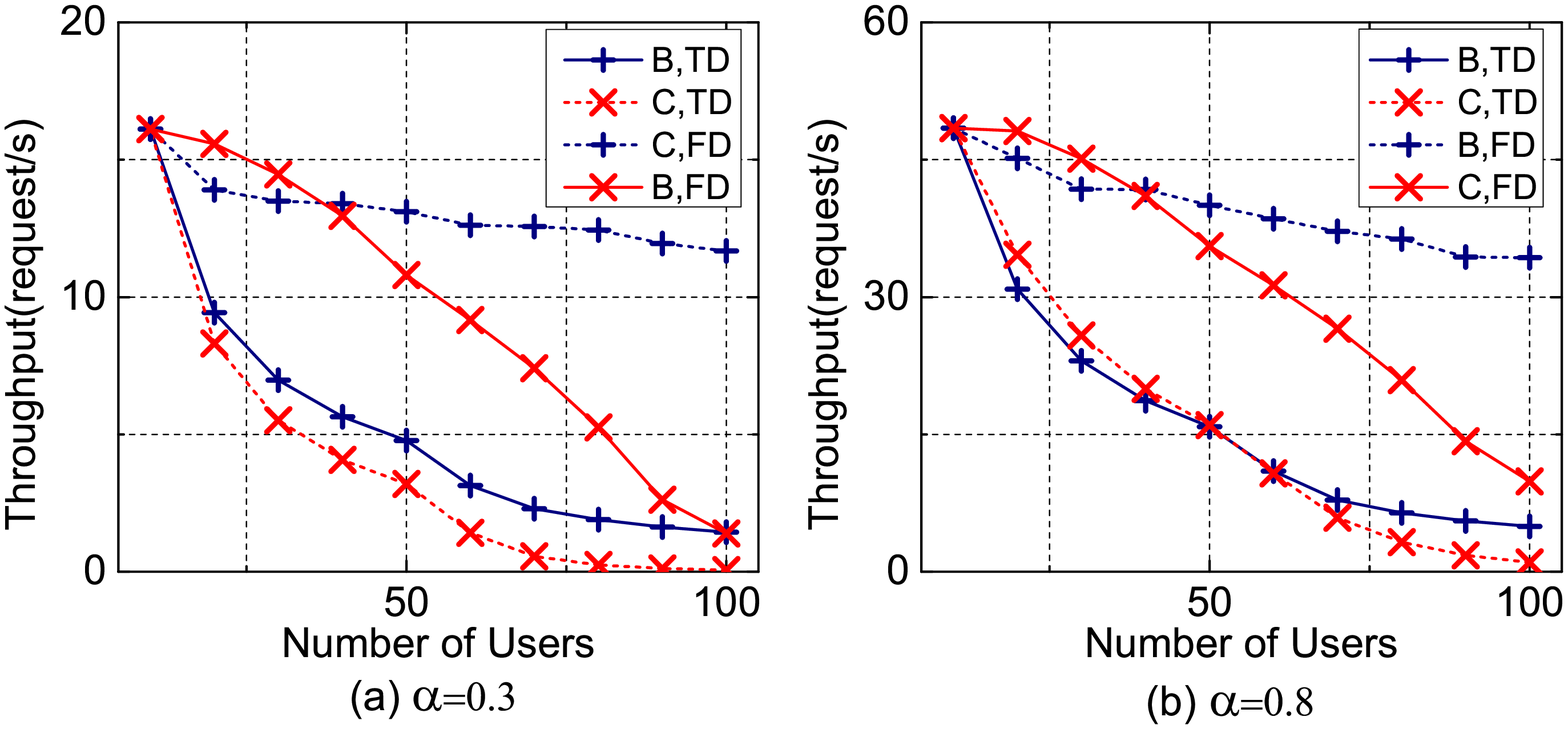}
\vspace{-2.7cm}
\caption{System Throughput versus number of users under two kinds of cache size.}
\label{fig:usernum}
\vspace{-0.2cm}
\end{figure}

Under the coded cache scheme, the system throughput becomes zero when the number of users is larger enough, regardless of any resource allocation strategies. Based on the coded cache scheme, most of  transmissions will be restricted by the user with worst channel gain. Since the worst channel gain will be zero when the number of users is infinity, the most parts of system throughput will be zero. Under the baseline scheme, the user with worst channel gain only restrict the throughput of itself, while the system throughput just decreases a little via efficient bandwidth allocation.

\section{Conclusion and Future Work}

In this paper, we apply the coded caching scheme into the wireless network with fading channel. We investigate the performance of coded caching scheme with resource allocation when consider the transmission mode in the wireless scenario and formulate a sub-optimal problem. Under the wireless environment with various performances of user channels, the system throughput yielded by the coded caching scheme is limited by the worse case users, and only shows the performance gain under large cache size and large system bandwidth. Moreover, with power or bandwidth allocation, the throughput performance of coded caching scheme under the FD transport mode is significantly better than that under the TD mode. Further simulations also show that its performance will decrease as the number of users becomes sufficiently large without considering the outage of users.

Recently, Wang~\cite{hetercache} proposes that the group coded delivery(GCD), i.e., divide users into group and operate coded caching scheme separately, can still guarantee the performance of traffic-volume under the heterogenous user cache sizes. In fact, when we adopt the GCD in our regime, it can effectively counteract the restrictions of the worst case user, since we can classify the users based on their channel gain and operate the resource allocation separately. In future work, we will investigate the deploy of GCD to further improve the performance of coded caching scheme in the wireless network.

\bibliographystyle{ieeetr}

 \end{document}